\let\ifarxiv=\iftrue     

\ifarxiv
\ifx\ifpdf\sdflkaj\let\ifpdf=\iffalse\fi
\documentclass[12pt,a4paper]{article}

\usepackage{amsmath,amssymb}
\usepackage[nosort]{cite}
\usepackage[hyperref,parsep]{collect}

\else
\documentclass[cits]{PoS}
\usepackage{amsmath,amssymb}

\ifpdf\fi

\fi

\makeatletter
\let\old@startsection=\@startsection
\renewcommand{\@startsection}[6]{\old@startsection{#1}{#2}{#3}{#4}{#5}{#6\mathversion{bold}}}
\makeatother

\makeatletter \@addtoreset{equation}{section} \makeatother

\makeatletter
\let\old@makecaption=\@makecaption
\def\@makecaption{\small\old@makecaption}
\makeatother

\let\oldPhi=\Phi
\let\oldPsi=\Psi
\let\oldGamma=\Gamma
\let\oldDelta=\Delta
\let\oldSigma=\Sigma
\let\oldLambda=\Lambda
\let\oldTheta=\Theta
\let\oldPi=\Pi
\renewcommand{\Phi}{\mathnormal{\oldPhi}}
\renewcommand{\Psi}{\mathnormal{\oldPsi}}
\renewcommand{\Gamma}{\mathnormal{\oldGamma}}
\renewcommand{\Sigma}{\mathnormal{\oldSigma}}
\renewcommand{\Delta}{\mathnormal{\oldDelta}}
\renewcommand{\Theta}{\mathnormal{\oldTheta}}
\renewcommand{\Lambda}{\mathnormal{\oldLambda}}
\renewcommand{\Pi}{\mathnormal{\oldPi}}

\newcommand{\gen}[1]{\mathfrak{#1}}
\newcommand{\genY}[1]{\widehat{\mathfrak{#1}}}
\newcommand{\smat}{\mathcal{S}}

\newcommand{\eip}{\mathcal{U}}

\newcommand{\superN}{\mathcal{N}}
\newcommand{\copro}{\oldDelta}
\newcommand{\struc}{f}
\newcommand{\kkform}{g}

\newcommand{\counit}{\varepsilon}
\newcommand{\antipode}{\mathrm{S}}

\newcommand{\Reals}{\mathbb{R}}

\ifx\genfrac\sdflkaj

\else

\fi
\newcommand{\sfrac}[2]{{\textstyle\frac{#1}{#2}}}
\newcommand{\half}{\sfrac{1}{2}}

\newcommand{\quarter}{\sfrac{1}{4}}

\newcommand{\lrbrk}[1]{\left(#1\right)}
\newcommand{\bigbrk}[1]{\bigl(#1\bigr)}

\newcommand{\bigcomm}[2]{\big[#1,#2\big]}
\newcommand{\comm}[2]{[#1,#2]}

\newcommand{\acomm}[2]{\{#1,#2\}}

\newcommand{\state}[1]{\mathopen{|}#1\mathclose{\rangle}}

\newcommand{\xp}[1]{x^{+}_{#1}}
\newcommand{\xm}[1]{x^{-}_{#1}}
\newcommand{\xpm}[1]{x^{\pm}_{#1}}

\newcommand{\alg}[1]{\mathfrak{#1}}
\newcommand{\grp}[1]{\mathrm{#1}}

\newcommand{\nn}{\nonumber}
\newcommand{\nln}{\nonumber\\}
\newcommand{\nl}[1][0pt]{\nonumber\\[#1]&\hspace{-4\arraycolsep}&\mathord{}}

\newcommand{\earel}[1]{\mathrel{}&\hspace{-2\arraycolsep}#1\hspace{-2\arraycolsep}&\mathrel{}}
\newcommand{\eq}{\earel{=}}
\newcommand{\beq}{\begin{equation}}
\newcommand{\eeq}{\end{equation}}

\def\[{\begin{equation}}
\def\]{\end{equation}}
\def\<{\begin{eqnarray}}
\def\>{\end{eqnarray}}

\makeatletter
\def\mr@ignsp#1 {\ifx\:#1\@empty\else #1\expandafter\mr@ignsp\fi}%
\newcommand{\multiref}[1]{\begingroup
\xdef\mr@no@sparg{\expandafter\mr@ignsp#1 \: }%
\def\mr@comma{}%
\@for\mr@refs:=\mr@no@sparg\do{\mr@comma\def\mr@comma{,}\ref{\mr@refs}}%
\endgroup}
\makeatother

\newcommand{\hypref}[2]{\ifx\href\asklfhas #2\else\href{#1}{#2}\fi}

\newcommand{\secref}[1]{Sec.~\multiref{#1}}

\newcommand{\tabref}[1]{Tab.~\multiref{#1}}

\renewcommand{\eqref}[1]{(\multiref{#1})}

\ifx\href\asklfhas\newcommand{\href}[2]{#2}\fi
\newcommand{\arxivno}[1]{\href{http://arxiv.org/abs/#1}{#1}}

\ifarxiv\else

\title{The S-Matrix of AdS/CFT and Yangian Symmetry}

\ShortTitle{The S-Matrix of AdS/CFT and Yangian Symmetry}

\author{\speaker{Niklas Beisert}\\
Max-Planck-Institut f\"ur Gravitationsphysik\\%
Albert-Einstein-Institut\\%
Am M\"uhlenberg 1, 14476 Potsdam, Germany\\
        E-mail: \email{nbeisert@aei.mpg.de}}

\abstract{We review the algebraic construction of the S-matrix of AdS/CFT. 
We also present its symmetry algebra which turns out to be a Yangian
of the centrally extended $\mathfrak{su}(2|2)$ superalgebra.}

\FullConference{Bethe Ansatz: 75 years later \\
		October 19-21 2006,\\
		Brussels, Belgium}

\fi

\begin{document}

\ifarxiv

\begin{flushright}\footnotesize
\texttt{\arxivno{arxiv:0704.0400}}\\
\texttt{AEI-2007-019}
\end{flushright}
\vspace{0cm}

\begin{center}%
{\Large\textbf{\mathversion{bold}%
The S-Matrix of AdS/CFT\\
and Yangian Symmetry}\par} \vspace{1cm}%

\textsc{Niklas Beisert}\vspace{5mm}%

\textit{Max-Planck-Institut f\"ur Gravitationsphysik\\%
Albert-Einstein-Institut\\%
Am M\"uhlenberg 1, 14476 Potsdam, Germany}\vspace{3mm}%

\texttt{nbeisert@aei.mpg.de}\par\vspace{1cm}

\textbf{Abstract}\vspace{7mm}

\begin{minipage}{12.7cm}
We review the algebraic construction of the S-matrix of AdS/CFT. 
We also present its symmetry algebra which turns out to be a Yangian
of the centrally extended $\mathfrak{su}(2|2)$ superalgebra.
\end{minipage}

\end{center}
\vspace{1cm}
\hrule height 0.75pt
\vspace{1cm}
\fi


\section{Introduction and Overview}

Bethe's ansatz \cite{Bethe:1931hc}
for solving a one-dimensional integrable model
was and remains a powerful tool in contemporary theoretical physics:
75 years ago it solved one of the first models of quantum mechanics,
the Heisenberg spin chain \cite{Heisenberg:1928aa};
today it provides exact solutions for the spectra of
certain gauge and string theories and thus helps us 
understand their duality \cite{Maldacena:1998re} better.
Since the discovery of integrable structures 
in planar $\superN=4$ supersymmetric gauge theory \cite{Minahan:2002ve,Beisert:2003tq,Beisert:2003yb}
and in planar IIB string theory on $AdS_5\times S^5$ \cite{Mandal:2002fs,Bena:2003wd}
the tools for computing and comparing the spectra of both models have evolved rapidly. 
We now have complete asymptotic Bethe equations \cite{Beisert:2005fw,Beisert:2006ez}
which interpolate smoothly between 
the perturbative regimes in gauge and string theory
and which agree with all available data.

In this note we will focus on the S-matrix \cite{Staudacher:2004tk}
in the excitation picture above a ferromagnetic ground state.
We start by reviewing the algebraic construction of the
S-matrix in \secref{sec:Algebra}. In \secref{sec:Yangian} we subsequently
show that this S-matrix has indeed a larger symmetry algebra: a Yangian.

\section{The Universal Enveloping Algebra $\grp{U}(\alg{su}(2|2)\ltimes \Reals^2)$}
\label{sec:Algebra}

In this section the results on 
the S-matrix of AdS/CFT shall be reviewed from an algebraic point of view.
The applicable symmetry is a central extension $\alg{h}$ 
of the Lie superalgebra $\alg{su}(2|2)$ which we consider first. 
We continue by presenting 
the Hopf algebra structure of its universal enveloping algebra
and its fundamental representation. 
Finally, we comment on the S-matrix and its dressing phase factor.

\paragraph{Lie Superalgebra.}

The symmetry in the excitation picture for light cone string theory 
on $AdS_5\times S^5$ and for single-trace local operators in $\superN=4$ 
supersymmetric gauge theory is given by two copies of the Lie superalgebra 
\cite{Beisert:2004ry,Beisert:2005tm}
\[\alg{h}:=\alg{su}(2|2)\ltimes\Reals^2=\alg{psu}(2|2)\ltimes\Reals^3.\]
It is a central extension of the
standard Lie superalgebras $\alg{su}(2|2)$ or $\alg{psu}(2|2)$, 
see \cite{Nahm:1977tg}.
It is generated by the
$\alg{su}(2)\times\alg{su}(2)$ generators $\gen{R}^{a}{}_{b}$,
$\gen{L}^{\alpha}{}_{\beta}$, the supercharges
$\gen{Q}^{\alpha}{}_{b}$, $\gen{S}^{a}{}_{\beta}$ and the central
charges $\gen{C}$, $\gen{P}$, $\gen{K}$. 
The Lie brackets of the $\alg{su}(2)$ generators 
take the standard form
\[\begin{array}[b]{rclcrcl}
\comm{\gen{R}^a{}_b}{\gen{R}^c{}_d}\eq
\delta^c_b\gen{R}^a{}_d-\delta^a_d\gen{R}^c{}_b,
&&
\comm{\gen{L}^\alpha{}_\beta}{\gen{L}^\gamma{}_\delta}\eq
\delta^\gamma_\beta\gen{L}^\alpha{}_\delta-\delta^\alpha_\delta\gen{L}^\gamma{}_\beta,
\\[3pt]
\comm{\gen{R}^a{}_b}{\gen{Q}^\gamma{}_d}\eq
-\delta^a_d\gen{Q}^\gamma{}_b+\half \delta^a_b\gen{Q}^\gamma{}_d,
&&
\comm{\gen{L}^\alpha{}_\beta}{\gen{Q}^\gamma{}_d}\eq
+\delta^\gamma_\beta\gen{Q}^\alpha{}_d-\half \delta^\alpha_\beta\gen{Q}^\gamma{}_d,
\\[3pt]
\comm{\gen{R}^a{}_b}{\gen{S}^c{}_\delta}\eq
+\delta^c_b\gen{S}^a{}_\delta-\half \delta^a_b\gen{S}^c{}_\delta,
&&
\comm{\gen{L}^\alpha{}_\beta}{\gen{S}^c{}_\delta}\eq
-\delta^\alpha_\delta\gen{S}^c{}_\beta+\half \delta^\alpha_\beta\gen{S}^c{}_\delta.
\end{array}\]
The Lie brackets of two supercharges yield
\<
\acomm{\gen{Q}^\alpha{}_b}{\gen{S}^c{}_\delta}\eq
\delta^c_b\gen{L}^\alpha{}_\delta +\delta^\alpha_\delta\gen{R}^c{}_b
+\delta^c_b\delta^\alpha_\delta\gen{C},
\nln
\acomm{\gen{Q}^{\alpha}{}_{b}}{\gen{Q}^{\gamma}{}_{d}}\eq
\varepsilon^{\alpha\gamma}\varepsilon_{bd}\gen{P}, 
\nln
\acomm{\gen{S}^{a}{}_{\beta}}{\gen{S}^{c}{}_{\delta}}\eq
\varepsilon^{ac}\varepsilon_{\beta\delta}\gen{K}.
\>
The remaining Lie brackets vanish.

Where appropriate, we shall use the
collective symbol $\gen{J}^A$ for 
the generators. 
The Lie brackets then take the standard form
\[\label{eq:Lie}
\comm{\gen{J}^A}{\gen{J}^B}=\struc^{AB}_C \gen{J}^C.
\]
For simplicity of notation, we shall pretend that all generators are bosonic;
the generalisation to fermionic generators by insertion of suitable signs 
and graded commutators is straightforward.

\paragraph{Hopf Algebra.}

\begin{table}\centering
\<
\copro\gen{R}^a{}_b\eq
\gen{R}^a{}_b\otimes 1
+1\otimes\gen{R}^a{}_b,
\nln
\copro\gen{L}^\alpha{}_\beta\eq
\gen{L}^\alpha{}_\beta\otimes 1
+1\otimes\gen{L}^\alpha{}_\beta,
\nln
\copro\gen{Q}^\alpha{}_b\eq
\gen{Q}^\alpha{}_b\otimes 1
+\eip^{+1}\otimes\gen{Q}^\alpha{}_b,
\nln
\copro\gen{S}^a{}_\beta\eq
\gen{S}^a{}_\beta\otimes 1
+\eip^{-1}\otimes\gen{S}^a{}_\beta,
\nln
\copro\gen{C}\eq
\gen{C}\otimes 1
+1\otimes\gen{C},
\nln
\copro\gen{P}\eq
\gen{P}\otimes 1
+\eip^{+2}\otimes\gen{P},
\nln
\copro\gen{K}\eq
\gen{K}\otimes 1
+\eip^{-2}\otimes\gen{K},
\nln
\copro\eip\eq
\eip\otimes \eip.
\nn
\>

\caption{The coproduct of the
braided universal enveloping algebra $\grp{U}(\alg{h})$.}
\label{tab:coprouni}
\end{table}

Next we consider the universal enveloping algebra 
$\grp{U}(\alg{h})$ of $\alg{h}$. The construction
of the product is standard, and one identifies
the Lie brackets \eqref{eq:Lie} with graded commutators. 
For the coproduct one can introduce a non-trivial braiding
\cite{Gomez:2006va,Plefka:2006ze} 
\[\label{eq:braid}
\copro\gen{J}^A=\gen{J}^A\otimes 1+\eip^{[A]}\otimes\gen{J}^A
\]
with some abelian%
\footnote{Curiously, we can include the supersymmetric
grading $(-1)^{\mathcal{F}}$ in the
generator $\eip$ to manually impose 
the correct statistics. 
This is helpful for an implementation 
within a computer algebra system. 
In this case $\eip$ would
anticommute with fermionic generators.}
generator $\eip$
(a priori unrelated to the algebra)
and the grading
\[
[\gen{R}]=[\gen{L}]=[\gen{C}]=0,\quad
[\gen{Q}]=+1,\quad
[\gen{S}]=-1,\quad
[\gen{P}]=+2,\quad
[\gen{K}]=-2.
\]
The coproduct is spelt out in \tabref{tab:coprouni} for
the individual generators.
The above grading is derived from the Cartan charge of the
$\alg{sl}(2)$ automorphism \cite{Nahm:1977tg} of the algebra $\alg{h}$
and therefore the coproduct is compatible with the algebra relations.

We should define the remaining structures of the Hopf algebra: 
the antipode $\antipode$ and the counit $\counit$ \cite{Gomez:2006va,Plefka:2006ze}.
The antipode is an anti-homomorphism which acts 
on the generators as
\[
\antipode(1)=1,
\quad
\antipode(\eip)=\eip^{-1},
\quad
\antipode(\gen{J}^A)=-\eip^{-[A]}\gen{J}^A.
\]
The counit acts non-trivially only on $1$ and $\eip$
\[
\counit(1)=\counit(\eip)=1,
\quad
\counit(\gen{J}^A)=0.
\]

\paragraph{Cocommutativity.}

This coproduct is in general not quasi-cocommutative
as can easily be seen by considering the
central charges $\gen{P}$, $\gen{K}$ in \tabref{tab:coprouni}.
To make it quasi-co\-commu\-ta\-tive we have to satisfy the constraints
\cite{Gomez:2006va}
\[\label{eq:PKconstr}
\gen{P}\otimes \bigbrk{1-\eip^{+2}}
=\bigbrk{1-\eip^{+2}}\otimes\gen{P},
\qquad
\gen{K}\otimes \bigbrk{1-\eip^{-2}}
=\bigbrk{1-\eip^{-2}}\otimes\gen{K}.
\]
They are solved by identifying the central charges
$\gen{P}$, $\gen{K}$ with the braiding factor
$\eip$ as follows \cite{Plefka:2006ze}
\[\label{eq:PKsol}
\gen{P}=g\alpha\bigbrk{1-\eip^{+2}},
\qquad
\gen{K}=g\alpha^{-1}\bigbrk{1-\eip^{-2}}.
\]
This leads to the following
quadratic constraint
\[
\gen{P}\gen{K}
-g\alpha^{-1}\gen{P}
-g\alpha\gen{K}
=0.
\]
It was furthermore shown
in \cite{Beisert:2006qh} that the 
coproduct is quasi-triangular,
at least at the level of central charges,
see also \cite{Beisert:2008tw}.

\paragraph{Fundamental Representation.}

The algebra $\alg{h}$ has a four-dimensional
representation \cite{Beisert:2005tm}
which we will call fundamental.
The corresponding multiplet has two bosonic states $\state{\phi^a}$ 
and two fer\-mi\-onic states $\state{\psi^\alpha}$.
The action of the two sets of $\alg{su}(2)$ generators 
has to be canonical
\<
\gen{R}^a{}_b\state{\phi^c}\eq\delta^c_b\state{\phi^a}
  -\half \delta^a_b\state{\phi^c},
\nln
\gen{L}^\alpha{}_\beta\state{\psi^\gamma}\eq\delta^\gamma_\beta\state{\psi^\alpha}
  -\half \delta^\alpha_\beta\state{\psi^\gamma}.
\>
The supersymmetry generators must also act in a manifestly
$\alg{su}(2)\times\alg{su}(2)$ covariant way
\<
\gen{Q}^\alpha{}_a\state{\phi^b}\eq a\,\delta^b_a\state{\psi^\alpha},\nln
\gen{Q}^\alpha{}_a\state{\psi^\beta}\eq b\,\varepsilon^{\alpha\beta}\varepsilon_{ab}\state{\phi^b},\nln
\gen{S}^a{}_\alpha\state{\phi^b}\eq c\,\varepsilon^{ab}\varepsilon_{\alpha\beta}\state{\psi^\beta},\nln
\gen{S}^a{}_\alpha\state{\psi^\beta}\eq d\,\delta^\beta_\alpha\state{\phi^a}.
\>
We can write the four parameters $a,b,c,d$ 
using the parameters $\xpm{}$, $\gamma$ and the constants $g$, $\alpha$ as
\[
a=\sqrt{g}\,\gamma,\quad
b=\sqrt{g}\,\frac{\alpha}{\gamma}\lrbrk{1-\frac{\xp{}}{\xm{}}},\quad
c=\sqrt{g}\,\frac{i\gamma}{\alpha \xp{}}\,,\quad
d=\sqrt{g}\,\frac{\xp{}}{i\gamma}\lrbrk{1-\frac{\xm{}}{\xp{}}}.
\]
The parameters $\xpm{}$ (together with $\gamma$) label
the representation and they must obey the constraint
\[\label{eq:xpmIdent}
\xp{}+\frac{1}{\xp{}}-\xm{}-\frac{1}{\xm{}}=\frac{i}{g}\,.
\]
The three central charges $\gen{C},\gen{P},\gen{K}$ and $\eip$ 
are represented by the values $C,P,K$ and $U$ which read
\[
C=\frac{1}{2}\,\frac{1+1/\xp{}\xm{}}{1-1/\xp{}\xm{}}\,,
\quad
P=g\alpha\lrbrk{1-\frac{\xp{}}{\xm{}}},
\quad
K=\frac{g}{\alpha}\lrbrk{1-\frac{\xm{}}{\xp{}}},
\quad
U=\sqrt{\frac{\xp{}}{\xm{}}}\,.
\]
They furthermore obey the quadratic relation $C^2-PK=\quarter$. 
Note that the corresponding quadratic combination of central charges
$\gen{C}^2-\gen{P}\gen{K}$ is singled out by being
invariant under the $\alg{sl}(2)$ external automorphism.

\paragraph{Fundamental S-Matrix.}

In \cite{Beisert:2005tm,Beisert:2006qh}
an S-matrix acting on the tensor product of
two fundamental representations was derived. 
It was constructed by imposing invariance under 
the algebra $\alg{h}$ 
\[\label{eq:Sinv}
[\copro \gen{J}^A,\smat]=0.
\]
We will not reproduce the result here, it is given
in \cite{Beisert:2006qh}.
Note that we have to fix the parameters 
$\xi=U=\sqrt{\xp{}/\xm{}}$
in order to make the action of the generators
compatible with the coproduct \eqref{eq:braid}.%
\footnote{This identification removes all
braiding factors from the S-matrix 
in \protect\cite{Beisert:2006qh} which will
thus satisfy the standard Yang-Baxter (matrix) equation,
see also \protect\cite{Beisert:2005tm,Arutyunov:2006yd,Martins:2007hb}.}

This S-matrix has several interesting properties.
Firstly, it is not of difference form; it cannot be written
as a function of the difference of some spectral parameters.
Secondly, the S-matrix could be determined uniquely up to one overall function 
merely by imposing a Lie-type symmetry \eqref{eq:Sinv} \cite{Beisert:2005tm}. 
This unusual fact is related to an unusual feature of
representation theory of the algebra $\alg{h}$:
The tensor product of two fundamental representations is irreducible 
in almost all cases \cite{Beisert:2006qh}.

Intriguingly this S-matrix is 
equivalent to Shastry{}'s R-matrix \cite{Shastry:1986bb}
of the one-dimen\-sio\-nal Hubbard model \cite{Hubbard:1963aa}.
Furthermore the Bethe equations \cite{Beisert:2005tm}
contain two copies of the Lieb-Wu equations
for the Hubbard model \cite{Lieb:1968aa}.
These observations of \cite{Beisert:2006qh} 
establish a link between an important model
of condensed matter physics and string theory
(complementary to the one in \cite{Rej:2005qt}).

Finally, let us note that one can derive (asymptotic) Bethe equations
from the S-matrix and thus confirm the conjecture in \cite{Beisert:2005fw}. 
So far this step has been performed in two different ways:
by means of the nested coordinate \cite{Beisert:2005tm} and the algebraic \cite{Martins:2007hb}
Bethe ansatz.

\paragraph{Phase Factor.}

The remaining overall phase factor of the S-matrix clearly cannot 
be determined by demanding invariance under $\alg{h}$. 
The phase factor was computed to some approximation from gauge theory \cite{Serban:2004jf}
and from string theory \cite{Arutyunov:2004vx,Beisert:2005cw,Hernandez:2006tk,Gromov:2007aq}.
The problem of an algebraically undetermined phase factor
is in fact generic. Usually one imposes a further crossing symmetry 
relation to obtain a constraint on it. 
Indeed the known string phase factor is consistent with crossing 
symmetry \cite{Janik:2006dc} as was shown in \cite{Arutyunov:2006iu}.
By substituting a suitable ansatz \cite{Beisert:2005wv} for the phase factor
into the crossing symmetry relation a conjecture for
the all-orders phase factor at strong coupling was made in \cite{Beisert:2006ib}.

A corresponding all-orders expansion at weak coupling 
was presented in \cite{Beisert:2006ez}.
The latter conjecture was obtained by a sort of 
analytic continuation in the perturbative order of the series. 
Let us illustrate this principle by means 
of a very simple example:
Consider the rational function $f(x)=1/(1-x)$. It has the following
expansions at $x=0$ and at $x=\infty$
\[
f(x)\stackrel{x\to0}{=}\sum_{n=0}^\infty a_n x^n,
\qquad
f(x)\stackrel{x\to\infty}{=}\sum_{n=1}^\infty b_n x^{-n}
\]
with $a_n=1$ and $b_n=-1$. When we consider $a_n$ and $b_n$ as 
analytic functions of the index, we can make the observation
(``reciprocity'')
\[\label{eq:loopcont}
a_n=-b_{-n}.
\]
Of course there are various ways in which the two functions $+1$ and $-1$ could 
be related, but the choice \eqref{eq:loopcont} 
appears to work for a surprisingly large class of functions!%
\footnote{Among other physical examples, we have identified circular Maldacena-Wilson loops
\cite{Erickson:2000af} and non-critical string theory \cite{Gross:1990ay}
where this reciprocity can be applied. Furthermore, summation by 
the Euler-MacLaurin formula (also known as zeta-function regularisation) is consistent with it. 
I thank Curt Callan, Marcos Mari\~no and Tristan McLoughlin for discussions of this principle.}
It was proved in \cite{Kotikov:2006ts}
that it does apply for the conjectured expansion of the phase factor.
Very useful integral expressions for the phase have recently appeared
in \cite{Belitsky:2007zp,Dorey:2007xn}. 
The analytic expression of the dressing phase can formally be 
obtained from the $\alg{psu}(2,2|4)$ Bethe equations 
\cite{Sakai:2007rk} (see however appendix D in \cite{Rej:2007vm}) in 
analogy to the covariant approach of 
\cite{Mann:2005ab,Rej:2005qt,Gromov:2006dh}. While this proposal may 
seem to be encouraging in general, it is at the same time strange 
from the Hopf algebra point of view to use an S-matrix which does not 
obey the crossing relation \cite{Sakai:2007rk}. This calls for 
further investigations.

Several tests of the phase have recently appeared, they are
based on four-loop unitary scattering methods \cite{Bern:2006ew}, 
numerical evaluation \cite{Benna:2006nd,Beccaria:2007tk},
analytic methods \cite{Benna:2006nd,Kotikov:2006ts,Alday:2007qf,Kostov:2007kx}
and on taking a certain highly non-trivial limit \cite{Maldacena:2006rv}.

\section{The Yangian $\grp{Y}(\alg{su}(2|2)\ltimes \Reals^2)$}
\label{sec:Yangian}

In the section we investigate Yangian symmetry \cite{Drinfeld:1985rx,Drinfeld:1986in}
for the above S-matrix.
We will start with a very brief review of Yangian 
symmetry for generic S-matrices 
(see \cite{Bernard:1993ya,MacKay:2004tc} for more extensive reviews),
and then we apply the framework to the S-matrix discussed above.

\paragraph{Yangians and S-Matrices.}

Typically the symmetries of rational S-matrices 
are of Yangian type.
The Yangian $\grp{Y}(\alg{g})$ of a Lie algebra $\alg{g}$ 
is a deformation of the universal enveloping algebra 
of half the affine extension of $\alg{g}$.

More plainly, it is generated by the $\alg{g}$-generators
$\gen{J}^A$ and the Yangian generators 
$\genY{J}^A$. Their commutators take the generic form
\<\label{eq:Yadj}
\comm{\gen{J}^A}{\gen{J}^B}\eq \struc^{AB}_C\gen{J}^C,
\nln
\comm{\gen{J}^A}{\genY{J}^B}\eq \struc^{AB}_C\genY{J}^C,
\>
and they should obey the Jacobi and Serre relations
\<\label{eq:Serre}
\bigcomm{\gen{J}^{[A}}{\comm{\gen{J}^B}{\gen{J}^{C]}}}
\eq 0,
\nln
\bigcomm{\gen{J}^{[A}}{\comm{\gen{J}^B}{\genY{J}^{C]}}}
\eq 0,
\nln
\bigcomm{\genY{J}^{[A}}{\comm{\genY{J}^B}{\gen{J}^{C]}}}
\eq \sfrac{1}{4}\hbar^2
\struc^{AG}_{D}\struc^{BH}_E\struc^{CK}_{F}\struc_{GHK}
\gen{J}^{\{D}\gen{J}^E\gen{J}^{F\}}.
\>
The symbol $\struc_{ABC}=\kkform_{AD}\kkform_{BE}\struc^{DE}_C$ represents
the structure constants $\struc^{AD}_C$ with two indices lowered
by means of the inverse of the Cartan-Killing forms $\kkform_{AD}$ and $\kkform_{BE}$.
The brackets $\{\phantom{x}\}$ and $[\phantom{x}]$ at the level of indices
imply total symmetrisation and anti-symmetrisation, respectively.
Finally, $\hbar$ is a scale parameter whose value plays no physical role.
The first two relations lead to a constraint on
the structure constants $\struc^{AB}_C$.
The third relation%
\footnote{For $\alg{g}=\alg{su}(2)$ it has to be replaced by a 
quartic relation.}
is a deformation of the Serre relation for
affine extensions of Lie algebras. 

The Yangian is a Hopf algebra and the coproduct takes 
the standard form
\<
\copro\gen{J}^A\eq\gen{J}^A\otimes 1+1\otimes\gen{J}^A,
\nln
\copro\genY{J}^A\eq\genY{J}^A\otimes 1+1\otimes\genY{J}^A+\half\hbar \struc^{A}_{BC}\gen{J}^B\otimes\gen{J}^C.
\>
where $\struc^{A}_{BC}=\kkform_{BD}\struc^{AD}_C$. 
The antipode $\antipode$ is defined by
\[
\antipode(\gen{J}^A)=-\gen{J}^A,
\quad
\antipode(\genY{J}^A)=-\genY{J}^A+\quarter\hbar\struc^A_{BC}\struc^{BC}_D\gen{J}^D,
\]
and the counit $\counit$ takes the standard form
\[
\counit(1)=1,\quad \counit(\gen{J}^A)=\counit(\genY{J}^A)=0.
\]

For the study of integrable systems, the evaluation representations 
of the Yangian are of special interest. For these
the action of the Yangian generators $\genY{J}^A$ 
is proportional to the Lie generators
\[
\genY{J}^A\state{u}=\hbar u \gen{J}^A\state{u}.
\]
Here $\state{u}$ is some state of the evaluation module
with spectral parameter $u$.
This Yangian representation is finite-dimensional if the
$\alg{g}$-representation is. One merely has to ensure
that the Serre relation \eqref{eq:Serre} is satisfied.
This is indeed not the case for all representations
of all Lie algebras.
The power of the Yangian symmetry lies in the fact that 
tensor products of evaluation representations
are typically irreducible 
(except for special values of their spectral parameters).
This allows for simple proofs (e.g.~for the Yang-Baxter relation)
by representation theory arguments.

Let us finally consider the connection to the S-matrix.
The S-matrix is a permutation operator; it acts by interchanging
two modules of the algebra
\[
\smat:\mathbb{V}_1\otimes\mathbb{V}_2\to\mathbb{V}_2\otimes\mathbb{V}_1.
\]
In particular, for the tensor product of two evaluation modules 
one has
\[
\smat\state{u_1,u_2}
\sim\state{u_2,u_1}.
\]
Invariance of the S-matrix under the Yangian means 
\[\label{eq:SYinv}
\comm{\copro\gen{J}^A}{\smat}=
\comm{\copro\genY{J}^A}{\smat}=0
\]
for all generators $\gen{J}^A$, $\genY{J}^A$.
The existence of such an S-matrix is equivalent to 
quasi-cocommutativity of $\grp{Y}(\alg{g})$.
Note that only the difference of 
spectral parameters appears
in the invariance condition:
We can write the action of the coproduct of Yangian generators 
on the evaluation module $\state{u_1,u_2}$ as
\[
\copro\genY{J}^A\simeq
(u_1-u_2)\gen{J}^A\otimes 1
+u_2\copro\gen{J}^A
+\hbar \struc^{A}_{BC}\gen{J}^B\otimes\gen{J}^C.
\]
Here the first equation in \eqref{eq:SYinv}
ensures that the term proportional
to $u_2$ drops out from the second equation.
Therefore the S-matrix typically depends 
on the difference $u_1-u_2$ of spectral parameters only.

\paragraph{Yangians in AdS/CFT.}

Yangian symmetries for planar AdS/CFT
have been investigated in \cite{Dolan:2003uh},
both for classical string theory and for gauge theory
at leading order,
see also 
\cite{Dolan:2004ps,Hatsuda:2004it,Dolan:2004ys}
Yangian symmetry also persists to higher perturbative orders in both models
\cite{Serban:2004jf,Agarwal:2004sz,Berkovits:2004xu,Zwiebel:2006cb,Beisert:2007jv}
and it is likely that it also exists at finite coupling.
This Yangian can be understood as a symmetry of the Hamiltonian on 
an infinite world sheet or as an expansion of the
full monodromy matrix. The Lie symmetry in 
this picture is $\alg{psu}(2,2|4)$ and the Yangian
would be $\grp{Y}(\alg{psu}(2,2|4))$.

Here we consider a different picture of well-separated excitations 
on a ferromagnetic ground state and of their scattering matrix. 
In this picture the Lie symmetry reduces to two copies
of $\alg{h}$ and the corresponding Yangian would be $\grp{Y}(\alg{h})$. 
Our Yangian should arise as a subalgebra of the full Yangian $\grp{Y}(\alg{psu}(2,2|4))$
when acting on asymptotic excitation states.

\paragraph{Hopf Algebra.}

Let us now consider $\grp{Y}(\alg{h})$.
We have already studied the universal enveloping algebra $\grp{U}(\alg{h})$. 
All we still need to do is to introduce one
generator $\genY{J}^A$ for each $\gen{J}^A$ 
obeying the relations \eqref{eq:Yadj,eq:Serre},
and define its coproduct, antipode as well as counit. 

In \eqref{eq:braid} we have defined a braided coproduct
for the universal enveloping algebra.
For consistency with the Serre relations, 
we also have to apply an analogous braiding to 
the standard Yangian coproduct 
\[\label{eq:YangCoproBraid}
\copro\genY{J}^A=
\genY{J}^A\otimes 1+\eip^{[A]}\otimes\genY{J}^A
+\hbar \struc^{A}_{BC}\gen{J}^B\eip^{[C]}\otimes\gen{J}^C.
\]
Note that lowering an index requires to use the
inverse Cartan-Killing form of the algebra.
In the case of $\alg{h}$ the Cartan-Killing form 
is degenerate and we need to extend the algebra by 
the $\alg{sl}(2)$ outer automorphism, 
see \cite{Beisert:2006qh}. Effectively, lowering
an index leads to an interchange 
of the automorphism generators
with the central charges.
We refrain from spelling out the
Cartan-Killing form or the modified structure constants.  
Instead we present the complete set of coproducts of
Yangian generators in \tabref{tab:copro},
where we also fix the value of $\hbar$.

\begin{table}\centering
\<
\copro\genY{R}^a{}_b\eq
\genY{R}^a{}_b\otimes 1
+1\otimes\genY{R}^a{}_b
\nl
+\half\gen{R}^a{}_c\otimes\gen{R}^c{}_b
-\half\gen{R}^c{}_b\otimes\gen{R}^a{}_c
\nl
-\half\gen{S}^a{}_\gamma\eip^{+1}\otimes\gen{Q}^\gamma{}_b
-\half\gen{Q}^\gamma{}_b\eip^{-1}\otimes\gen{S}^a{}_\gamma
\nl\qquad
+\quarter\delta^a_b\,\gen{S}^d{}_\gamma\eip^{+1}\otimes\gen{Q}^\gamma{}_d
+\quarter\delta^a_b\,\gen{Q}^\gamma{}_d\eip^{-1}\otimes\gen{S}^d{}_\gamma,
\nln
\copro\genY{L}^\alpha{}_\beta\eq
\genY{L}^\alpha{}_\beta\otimes 1
+1\otimes\genY{L}^\alpha{}_\beta
\nl
-\half\gen{L}^\alpha{}_\gamma\otimes\gen{L}^\gamma{}_\beta
+\half\gen{L}^\gamma{}_\beta\otimes\gen{L}^\alpha{}_\gamma
\nl
+\half\gen{Q}^\alpha{}_c\eip^{-1}\otimes\gen{S}^c{}_\beta
+\half\gen{S}^c{}_\beta\eip^{+1}\otimes\gen{Q}^\alpha{}_c
\nl\qquad
-\quarter\delta^\alpha_\beta\,\gen{Q}^\delta{}_c\eip^{-1}\otimes\gen{S}^c{}_\delta
-\quarter\delta^\alpha_\beta\,\gen{S}^c{}_\delta\eip^{+1}\otimes\gen{Q}^\delta{}_c,
\nln
\copro\genY{Q}^\alpha{}_b\eq
\genY{Q}^\alpha{}_b\otimes 1
+\eip^{+1}\otimes\genY{Q}^\alpha{}_b
\nl
-\half\gen{L}^\alpha{}_\gamma\eip^{+1}\otimes\gen{Q}^\gamma{}_b
+\half\gen{Q}^\gamma{}_b\otimes\gen{L}^\alpha{}_\gamma
\nl
-\half\gen{R}^c{}_b\eip^{+1}\otimes\gen{Q}^\alpha{}_c
+\half\gen{Q}^\alpha{}_c\otimes\gen{R}^c{}_b
\nl
-\half\gen{C}\eip^{+1}\otimes\gen{Q}^\alpha{}_b
+\half\gen{Q}^\alpha{}_b\otimes\gen{C}
\nl
+\half\varepsilon^{\alpha\gamma}\varepsilon_{bd}\gen{P}\eip^{-1}\otimes\gen{S}^d{}_\gamma
-\half\varepsilon^{\alpha\gamma}\varepsilon_{bd}\gen{S}^d{}_\gamma\eip^{+2}\otimes\gen{P},
\nln
\copro\genY{S}^a{}_\beta\eq
\genY{S}^a{}_\beta\otimes 1
+\eip^{-1}\otimes\genY{S}^a{}_\beta
\nl
+\half\gen{R}^a{}_c\eip^{-1}\otimes\gen{S}^c{}_\beta
-\half\gen{S}^c{}_\beta\otimes\gen{R}^a{}_c
\nl
+\half\gen{L}^\gamma{}_\beta\eip^{-1}\otimes\gen{S}^a{}_\gamma
-\half\gen{S}^a{}_\gamma\otimes\gen{L}^\gamma{}_\beta
\nl
+\half\gen{C}\eip^{-1}\otimes\gen{S}^a{}_\beta
-\half\gen{S}^a{}_\beta\otimes\gen{C}
\nl
-\half\varepsilon^{ac}\varepsilon_{\beta\delta}\gen{K}\eip^{+1}\otimes\gen{Q}^\delta{}_c
+\half\varepsilon^{ac}\varepsilon_{\beta\delta}\gen{Q}^\delta{}_c\eip^{-2}\otimes\gen{K},
\nln
\copro\genY{C}\eq
\genY{C}\otimes 1
+1\otimes\genY{C}
\nl
+\half\gen{P}\eip^{-2}\otimes\gen{K}
-\half\gen{K}\eip^{+2}\otimes\gen{P},
\nln
\copro\genY{P}\eq
\genY{P}\otimes 1
+\eip^{+2}\otimes\genY{P}
\nl
-\gen{C}\eip^{+2}\otimes\gen{P}
+\gen{P}\otimes\gen{C},
\nln
\copro\genY{K}\eq
\genY{K}\otimes 1
+\eip^{-2}\otimes\genY{K}
\nl
+\gen{C}\eip^{-2}\otimes\gen{K}
-\gen{K}\otimes\gen{C}.
\nn
\>
\caption{The coproduct of the Yangian generators in $\grp{Y}(\alg{h})$.}
\label{tab:copro}
\end{table}

For the sake of completeness we state the antipode%
\footnote{Note that $\struc^A_{BC}\struc^{BC}_D=0$ here,
so there is no contribution from the Lie generators.}
and the counit
\[
\antipode(\genY{J}^A)=-\eip^{-[A]}\genY{J}^A,
\qquad
\counit(\genY{J}^A)=0.
\]
%

\paragraph{Cocommutativity.}

An important question is if this coproduct can be quasi-co\-com\-mu\-ta\-tive.%
\footnote{The braiding factors in
\protect\eqref{eq:YangCoproBraid} turn out to be very important for the Yangian.
It can easily be seen that without them the coproduct cannot
be quasi-cocommutative. This is in contradistinction 
to the universal enveloping algebra where the
braided as well as the unbraided coproduct 
are quasi-cocommutative.}
A first step is to consider the central generators
$\genY{C}$, $\genY{P}$, $\genY{K}$.
For that purpose it is favourable to choose
suitable combinations
\<\label{eq:CYangNew}
\genY{C}'\eq\genY{C}+\half g\alpha^{-1}\gen{P}-\half g\alpha\gen{K},
\nln
\genY{P}'\eq\genY{P}+\gen{C}\bigbrk{\gen{P}-2g\alpha},
\nln
\genY{K}'\eq\genY{K}-\gen{C}\bigbrk{\gen{K}-2g\alpha^{-1}},
\>
for whom the coproduct almost trivialises
\<\label{eq:CYangCopro}
\copro\genY{C}'\eq
\genY{C}'\otimes 1
+1\otimes\genY{C}',
\nln
\copro\genY{P}'\eq
\genY{P}'\otimes 1
+\eip^{+2}\otimes\genY{P}',
\nln
\copro\genY{K}'\eq
\genY{K}'\otimes 1
+\eip^{-2}\otimes\genY{K}'.
\>
The combination $\genY{C}'$ is already cocommutative,
and in order to make the generators $\genY{P}'$, $\genY{K}'$ cocommutative
we have to set as above in \eqref{eq:PKconstr,eq:PKsol}
\[\label{eq:YPKSol}
\genY{P}'=igu_{\gen{P}}\gen{P},
\qquad
\genY{K}'=igu_{\gen{K}}\gen{K}
\]
with two universal constants $u_{\gen{P}}$ and $u_{\gen{K}}$.
With this choice, $\genY{C}$, $\genY{P}$, $\genY{K}$
also become cocommutative because 
they differ from $\genY{C}'$, $\genY{P}'$, $\genY{K}'$
only by central elements.

\ifarxiv\else\newpage\fi
\paragraph{Fundamental Evaluation Representation.}

For the fundamental evaluation representation 
we make the ansatz%
\footnote{We believe, but we have not verified that this is compatible
with the Serre relations \protect\eqref{eq:Serre}.}
\[
\genY{J}^A\state{\mathcal{X}}=ig(u+u_0) \gen{J}^A\state{\mathcal{X}}.
\]
By comparison with \eqref{eq:CYangNew,eq:YPKSol} we can infer that 
$u$ has to be related to the parameters of the fundamental representation
by
\[\label{eq:relxpmu}
u=\xp{}+\frac{1}{\xp{}}-\frac{i}{2g}=\xm{}+\frac{1}{\xm{}}+\frac{i}{2g}
=\half(\xp{}+\xm{})(1+1/\xp{}\xm{})\,.
\]
Furthermore $u_{\gen{P}}$ and $u_{\gen{K}}$ in \eqref{eq:YPKSol}
have to both coincide with the universal constant
$u_0=u_{\gen{P}}=u_{\gen{K}}$.%
\footnote{It is conceivable that a further consistency requirement fixes the value of $u_0$,
presumably to zero.}

As an aside we state the eigenvalue of 
the quadratic combination
\[
C\widehat{C}-\half P\widehat{K}-\half K\widehat{P}
=\quarter ig(u+u_0).
\]

\paragraph{Fundamental S-Matrix.}

Using the coproducts in \tabref{tab:copro}
we have confirmed that the S-matrix 
is also invariant under 
all of the Yangian generators
\[
[\copro \genY{J}^A,\smat]=0.
\]
We have used a computer algebra system
to evaluate the action of the Yangian generators and the S-matrix.%
\footnote{We have also confirmed the invariance of the
singlet state found in \protect\cite{Beisert:2005tm}.}
To show invariance requires heavy use of the identity \eqref{eq:xpmIdent}. 
Superficially it is very surprising to
find all these additional symmetries of the S-matrix. 
The deeper reason however should be that the coproduct is quasi-cocommutative.
We have thus proved quasi-cocommutativity when acting on fundamental representations.

It is interesting to see that the S-matrix is based on standard evaluation 
representations of the Yangian. Nevertheless, it is not a function of the
difference of spectral parameters. This unusual property traces
back to the link between the spectral parameter $u$ and
the $\alg{h}$-representation parameters $\xpm{}$ in \eqref{eq:relxpmu}.
The latter is again related to the braiding in the coproduct
\eqref{eq:YangCoproBraid}.

As our S-matrix is equivalent \cite{Beisert:2006qh}
to Shastry{}'s R-matrix,
our Yangian is presumably an extension of the 
$\alg{su}(2)\times\alg{su}(2)$ Yangian
symmetry of the Hubbard model found in \cite{Uglov:1993jy}.

\section{Conclusions and Outlook}

In this note we have reviewed the construction of the S-matrix
with centrally extended $\alg{su}(2|2)$ symmetry
that appears in the context of the planar AdS/CFT correspondence
and the one-dimensio\-nal Hubbard model.
We have furthermore shown that the S-matrix has an additional Yangian symmetry 
whose Hopf algebra structure we have presented. 
This Yangian is not quite a standard Yangian, but its
coproduct needs to be braided in order to be quasi-cocommutative. 
This fact is intimately 
related to the existence of a triplet of central charges
with non-trivial coproduct 
and leads to the wealth of unusual features of the S-matrix.

In connection to the Yangian there are many points left to be clarified. 
Most importantly the representation theory needs to be understood. 
Which representations of $\alg{h}$ lift to evaluation representations
of $\grp{Y}(\alg{h})$? At what values of the spectral parameters do 
their tensor products become reducible?
This information could be used to prove that the 
coproduct is quasi-cocommutative. Also the 
Yang-Baxter equation for the S-matrix should follow straightforwardly.
It might also give some further understanding of bound states
\cite{Dorey:2006dq,Chen:2006gq}. 

Then it would be highly desirable to construct a universal R-matrix 
for this Yangian and show that it is quasi-triangular. 
This would put large parts of the integrable structure for
arbitrary representations of this algebra on solid ground
much like for the case of generic simple Lie algebras.

Some further interesting questions include:
Is this Yangian the unique quasi-co\-commu\-ta\-tive
Hopf algebra based on $\alg{h}$? Does the double Yangian \cite{Drinfeld:1986in}
exist and what is its structure?
Can the $\alg{sl}(2)$ automorphism of the algebra 
be included at the Yangian level such that the 
coproduct is quasi-cocommutative? 
What would the representations be in this case?

\paragraph{Acknowledgements.}

\ifarxiv
I am grateful to
\else
I thank
\fi
C.~Callan, D.~Erkal, A.~Kleinschmidt, P.~Koroteev, N.~MacKay,
M.~Mari\~no, T.~McLoughlin, J.~Plefka, F.~Spill
and B.~Zwiebel for interesting discussions.

\bibliography{su22yang}
\ifarxiv
\bibliographystyle{nbshort}
\else
\bibliographystyle{nbshortpos}
\fi

\end{document}